\documentclass[published]{epl}
\issue{60}{1}{2002}{120}{1 October 2002}
\recff{11 April 2002}{15 July 2002}
\usepackage{amssymb}
\usepackage[latin1]{inputenc}

\title{Disorder-induced enhancement of the persistent current for
       strongly interacting electrons in one-dimensional rings}
\shorttitle{Enhancement of persistent currents by disorder}
\author{Elise Gambetti-Césare\inst{1} \and 
Dietmar Weinmann\inst{1}\thanks{E-mail: \email{weinmann@ipcms.u-strasbg.fr}}
\and Rodolfo A.\ Jalabert\inst{1,2} \and Philipp Brune\inst{3}}
\shortauthor{E. Gambetti-Césare \etal}
\institute{
\inst{1} Institut de Physique et Chimie des Matériaux de Strasbourg,
UMR 7504 (CNRS-ULP), 23 rue du Loess, 67037 Strasbourg Cedex, France\\
\inst{2} Institute for Theoretical Physics, University of California, 
Santa Barbara, CA 93106, USA\\ 
\inst{3} Institut für Physik, Universität Augsburg, 86135 Augsburg, Germany
}
\begin{document}

\maketitle

\begin{abstract}
We show that disorder increases the persistent current of 
a half-filled one-dimensional Hubbard-Anderson ring at strong interaction. 
This unexpected effect results from a perturbative expansion starting
from the strongly interacting Mott insulator ground state. 
The analytical result is confirmed and extended by numerical 
calculations.
\end{abstract}

The interplay between disorder and electron-electron interactions is a
challenging problem, at the heart of the understanding of recent
important experimental findings. In particular, the measured
persistent currents \cite{pc_exp} in metallic mesoscopic rings are
much larger than the theoretical prediction for noninteracting
electrons \cite{pc_theo}, and the metallic phase appearing in
two-dimensional electron gases at low-density
\cite{kravchenko94,abrahams01} is at odds with the scaling theory 
of localization \cite{abrahams79} for noninteracting disordered systems.

Given the complexity of treating disorder and interactions on the same
footing, only limiting cases or simplified models have been solved. 
In such cases, the emerging conclusion is that both, disorder and a
repulsive interaction suppress the mobility and persistent currents. 
This suppression is expected to be even more important in mesoscopic 
systems of reduced dimensionality, where the role of correlations is 
enhanced by the poor screening.

These intuitive conclusions have been challenged in few-particle models, 
where it has been shown that a repulsive interaction, in the presence
of strong disorder, may result in an enhancement of the
electron mobility \cite{shepelyansky94, benenti99}. Such
delocalization effects are very dependent on the model and its
particular dimensionality, and their connection with the
experimentally relevant cases is still a matter of debate. 

In this work we show that disorder may also have the unexpected effect
of favoring the zero-temperature persistent currents, in the case of
strongly interacting electrons in one-dimensional half-filled Hubbard
rings. This result is established from a perturbative expansion starting 
from the strongly interacting limit, and numerical calculations based
on the Density Matrix Renormalization Group algorithm 
\cite{white93,peschel99}. 
Within a different system, namely a two-dimensional disordered Hubbard
model, numerical indications of such a behavior for the finite
temperature conductivity have recently been reported
\cite{denteneer01}, and related with the disorder induced breaking of
the particle-hole symmetry.

A large variety of analytical and numerical tools have been developed
and used for the study of interacting fermions in disordered
one-dimensional systems. In particular, for spinless fermions, it has
been numerically shown that repulsive interactions suppress the
persistent currents, provided that the disorder is not too strong
\cite{bouzerar94,kato94}. Only for very strong disorder, an
enhancement due to repulsive interactions occurs 
\cite{abraham93,schmitteckert98,weinmann01}. In this case, there is an
important difference between the behavior of individual samples (exhibiting
pronounced current peaks) and ensemble averages (showing a broad
region of small enhancement for moderate values of the
interaction). Once the spin is included in the description
(Hubbard-Anderson model), renormalization group \cite{giamarchi95} and
numerical approaches \cite{roemer95,mori96} show that, outside half
filling, repulsive interactions enhance the persistent currents in
weakly disordered systems. Our work extends these results towards half
filling and strong disorder, showing that the interaction induced
increase of the persistent current is much more important than in the
spinless case.  

We write the one-dimensional Hubbard-Anderson Hamiltonian as 
$H=H_0+H_1$ with
\begin{equation}\label{hamiltonian}
H_0= U\sum_{i=1}^{M}n_{i,\uparrow}n_{i,\downarrow} +
W\sum_{i=1}^{M}\sum\limits_{\sigma}v_in_{i,\sigma}
\quad \mbox{and} \quad 
H_1=-t\sum_{i=1}^{M}\sum_{\sigma}
\left( c_{i,\sigma}^{\dagger}c_{i-1,\sigma}^{\phantom{\dagger}}
+c^{\dagger}_{i-1,\sigma}c_{i,\sigma}^{\phantom{\dagger}}\right).
\end{equation}
$H_0$ accounts for interaction and disorder energy, and $H_1$ 
for kinetic energy. The operator $c_{i,\sigma}^{\phantom{\dagger}}$ 
($c_{i,\sigma}^{\dagger}$) destroys (creates) a particle with spin 
$\sigma$ on site $i$,
$n_{i,\sigma}=c_{i,\sigma}^{\dagger}c_{i,\sigma}^{\phantom{\dagger}}$ 
is the occupation number operator, and the hopping element $t=1$ sets 
the energy scale. The $v_i$ are independent random variables,
uniformly distributed within $[-1/2;1/2]$. $W$ and $U$ denote disorder
and interaction strengths, respectively. We concentrate on chains with an
even number of sites $M$ and $N=M$ electrons (half filling). 
The ring topology results from the boundary condition 
$c_0\equiv c_M\exp{i\Phi}$, where the phase $\Phi=2\pi\phi/\Phi_0$ 
accounts for a magnetic flux $\phi$ threading the ring, and
$\Phi_0=hc/e$ is the flux quantum.

The flux dependence of the many-body ground state energy $E(\Phi)$ is 
periodic with period $2\pi$. In localized systems, the persistent
current at zero temperature 
$I=-\partial E_0/\partial\Phi\approx (\Delta E/2)\sin\Phi$ can be 
characterized by the phase sensitivity $\Delta E=E(0)-E(\pi)$,
measuring the difference between the ground-state energies at 
periodic $(\Phi=0)$ and anti-periodic $(\Phi=\pi)$ boundary
conditions. We shall concentrate in the sequel on two related 
quantities: $\Delta E$ and the stiffness $D=M |\Delta E|/2$, which can be 
linked to the conductivity \cite{kohn} and the conductance of the 
system \cite{sushkov01}.

At half filling, strong repulsive interactions lead to a Mott-Hubbard 
transition and thereby to a strong suppression of the persistent
current \cite{abraham93}. For $U\gg t$, the kinetic energy part $H_1$
represents a small perturbation to the dominating term $H_0$ of the 
Hamiltonian (\ref{hamiltonian}), and it is possible to expand the 
ground-state energy in terms of $t/U$ 
\cite{tsiper97,tsiper98,selva00,weinmann01}. Since $H$ is symmetric
with respect to spin rotation and conserves the projection $S_z$ of
the total spin $S$ onto the $z$-axis, we choose to work in the subspace 
which is characterized by $S_z=0$, without restricting $S$. 
Within this subspace, the $N$-body eigenstates $|\psi_{\alpha}\rangle$
of $H_0$ are given by products of on-site states, and completely
specified by the functions $i_k^{\uparrow (\downarrow)} (\alpha)$ that
select the sites which are occupied by the
$N/2~\uparrow\!\!(\downarrow)$-electrons ($k=1,2,....N/2$) in the
configuration $\alpha$. They read 
\begin{equation}\label{state}
|\psi_{\alpha}\rangle =
\left(\prod_{k=1}^{N/2}c_{i^{\uparrow}_k(\alpha),\uparrow}^{\dagger}\right)
\left(\prod_{k=1}^{N/2}c_{i^{\downarrow}_k(\alpha),\downarrow}^{\dagger}\right)
|0\rangle \, . 
\end{equation}
The corresponding many-body energies are
$E_{\alpha}=E_{\alpha}^{\rm W}+E_{\alpha}^{\rm U}$, with the disorder 
contributions
$E_{\alpha}^{\rm W}=W\sum_k \left(v_{i^{\uparrow}_k(\alpha)}
+v_{i^{\downarrow}_k(\alpha)}\right)$, and the
interaction energies $E_{\alpha}^{\rm U}=U \epsilon_{\alpha}$. 
We note $\epsilon_{\alpha}$ the number of doubly occupied sites. 
For $U \gg W,t$ there is a gap separating the states without doubly 
occupied sites (subspace $\cal S$) from those having at least one 
doubly-occupied site (subspace $\cal D$). The $\cal S$-states are 
Mott-insulator configurations ($\epsilon_\alpha=0$) and all have the 
same lowest energy $E_0^{\rm W}=W\sum_i v_i$, independent of their
spin structure.
This degeneracy is broken by the hopping part  $H_1$ of the Hamiltonian.
Restricting ourselves to second order, we obtain an antiferromagnetic 
Heisenberg Hamiltonian as the effective model within the subspace 
$\cal S$ \cite{auerbach94}. Whenever neighbouring sites ($i$ and
$i-1$) have opposite spins, virtual hoppings 
$i \rightarrow i-1 \rightarrow i$ and $i-1 \rightarrow i \rightarrow i-1$ 
lead to a flux-independent coupling 
$-2t^2/[U(1-(v_i-v_{i-1})^2W^2/U^2)]$ between the two states that differ 
only in the orientation of the $i$ and $i-1$ spins. 

According to a theorem by Marshall, the non-degenerate ground state of
the antiferromagnetic Heisenberg Hamiltonian (with energy $E_0$) can
be written as $|\chi_0\rangle = \sum_{\beta \in {\cal S}} 
f_{\beta} |\psi_{\beta}\rangle$, where the weights $f_\beta$ can be
chosen real and positive with $0<f_\beta<1$ \cite{auerbach94}. The
effective model also contains diagonal matrix elements which are given
by the sum of the above mentioned terms over all pairs of adjacent
sites with opposite spin. The two configurations with alternating spins 
($i^{\uparrow(\downarrow)}_k=2k$ and $i^{\downarrow(\uparrow)}_k=2k-1$)
have the lowest diagonal elements matrix elements and therefore the
largest weights $f_\beta$.

We calculate the leading contribution to the phase sensitivity 
$\Delta E$, using the difference of
the higher order corrections for the ground-state energies at 
periodic ($\Phi=0$) and anti-periodic $(\Phi=\pi)$ boundary
conditions. The leading flux-dependent corrections to the 
ground-state energy appear in $M^{\rm th}$ order \cite{selva00} as
\begin{equation}\label{energyperturb}
E^{(M)} = 
\sum_{\gamma_1,\gamma_2,\dots ,\gamma_{M-1}}\,
\sum_{\beta,\beta'}f_\beta f_{\beta'}\frac{\langle \psi_\beta|H_1|\psi_{\gamma_1}\rangle\langle\psi_{\gamma_1}|
\cdots
|\psi_{\gamma_{M-1}}\rangle\langle\psi_{\gamma_{M-1}}|H_1|\psi_{\beta'}\rangle}
{(E_0-E_{\gamma_1})(E_0-E_{\gamma_2})\cdots (E_0-E_{\gamma_{M-1}})} \, .
\end{equation}
$\beta$ and $\beta'$ run over all the $\cal S$-states and  
the $\gamma$'s over the intermediate $\cal D$-states $|\psi_{\gamma} \rangle$. 
The numerator of the terms in (\ref{energyperturb}) is composed of 
matrix elements
$\langle\psi_{\gamma_l}|H_1|\psi_{\gamma_{l+1}}\rangle$ of the hopping
part of the Hamiltonian, thus non-zero contributions to $E^{(M)}$
arise only when all subsequent virtual 
states $|\psi_\gamma\rangle$ can be connected by one-particle
hopping processes \cite{selva00}. The resulting correction to 
the ground-state energy $E^{(M)}$ can be expressed using sums 
over sequences ${\bf A}^{(\beta,\beta')}$ of one-particle hopping
processes, starting at $|\psi_{\beta'} \rangle$ and ending at 
$|\psi_{\beta} \rangle$. 

The dependence of the corrections (\ref{energyperturb}) on 
the boundary condition enters {\em via}\/ the hopping terms
between sites $1\leftrightarrow M$ whose sign is reversed for
anti-periodic boundary conditions.
For periodic boundary conditions, all non-zero hopping elements 
are $-t$ and the numerators in (\ref{energyperturb}) are given by 
$(-t)^n {\rm sign}P_{\uparrow}({\bf A})\, {\rm
  sign}P_{\downarrow}({\bf A})$, 
with $P_{\uparrow(\downarrow)}({\bf A})$ being the 
permutation of the positions for $\uparrow\!\!(\downarrow)$ electrons
on the ring, resulting from the sequence ${\bf A}$. 
For anti-periodic boundary conditions, an additional sign 
$(-1)^{h_{\rm b}}$ appears, where $h_{\rm b}$ is the number of hops 
across the boundary $1\leftrightarrow M$ contained in ${\bf A}$. 

\begin{figure}[tb]
\includegraphics*[height=5cm,angle=0]{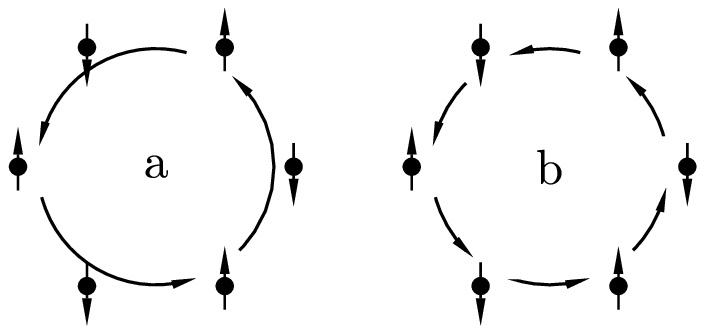}\hfill
\includegraphics*[height=5cm,angle=0]{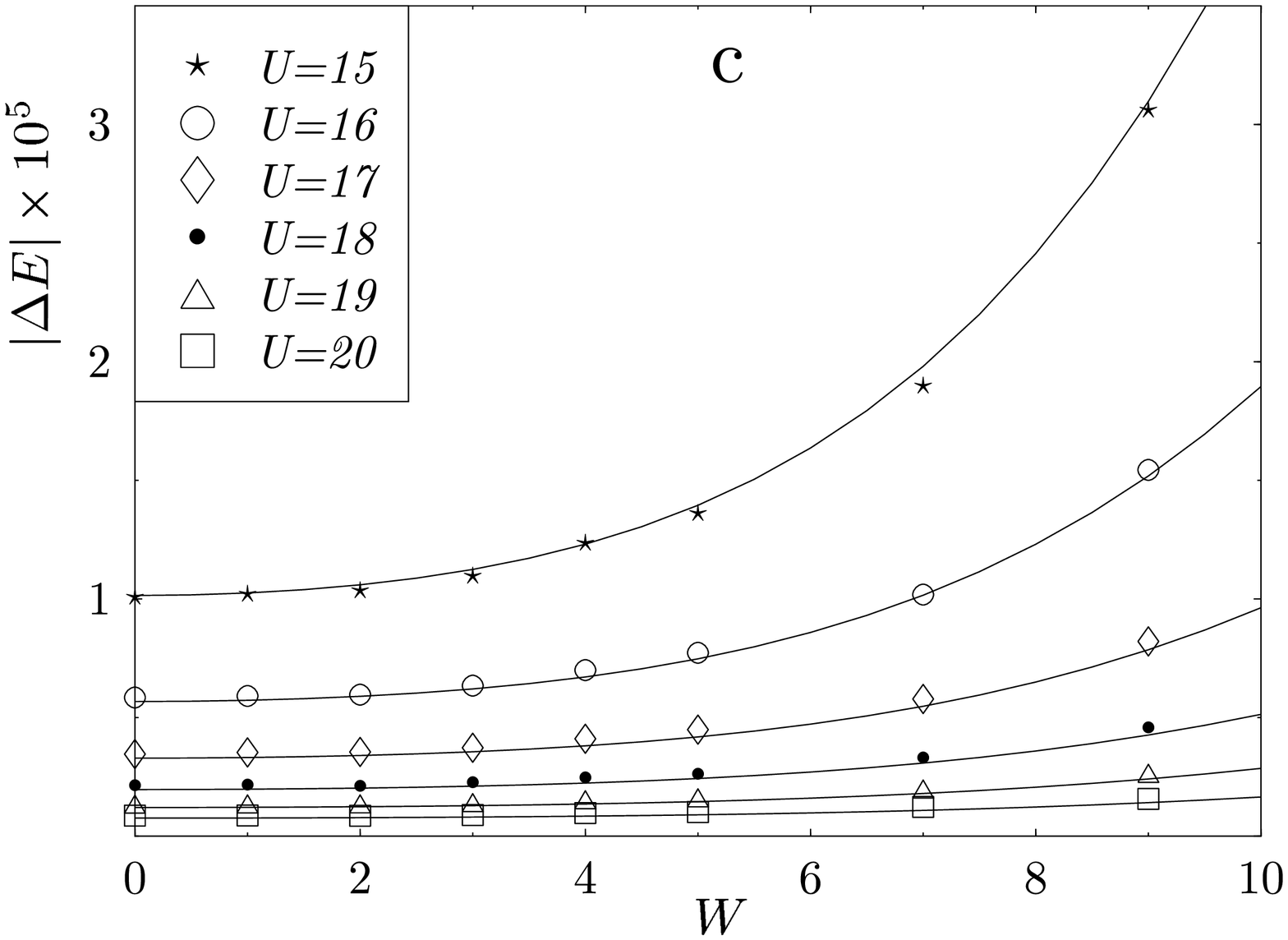}
\caption{\label{hops} 
(a,b): 
   Schematic illustration of hopping sequences of order $n=M$ 
   contributing to the phase sensitivity (\ref{DeltaE}), for the example
   $M=6$.
(c):
   The disorder dependence of $|\Delta E|$ for an individual sample with
   $M=N=10$. The data (symbols) are compared to the result of our perturbation
   theory (lines). 
}
\end{figure}
Only sequences ${\bf A}$ with odd $h_{\rm b}$ yield energy corrections
whose sign depends on the boundary condition. They are the only
contributions to the phase sensitivity $\Delta E$. The lowest
order sequences having odd $h_{\rm b}$ (and connecting one
ground-state component $|\psi_\beta\rangle$ from $\cal S$ to another
one) are those which contain exactly one hop between 
each given pair of neighbouring sites. All hops must be in the same 
direction along the ring and the net effect is the transfer of one 
particle around the ring. This is why we work in order $M$, which gives
the lowest order contributions with odd $h_{\rm b}=1$ \cite{excitedS}.
Typical sequences contributing to 
$\Delta E^{(M)}$ are shown in Fig.~\ref{hops}. Diagonal
contributions ($\beta=\beta'$) occur due to processes of type (a):
only electrons with a given spin move while the electrons with the
opposite spin orientation remain on their initial site.
Non-diagonal contributions ($\beta\neq\beta'$) can also appear. An
example for which each of the electrons moves by one site is depicted
in Fig.~\ref{hops}(b). 
Thus, the contributions to $\Delta E^{(M)}$ are given by those 
hopping sequences which contain either $M$ forward or $M$ backward
hops. Using (\ref{energyperturb}), we can express 
\begin{equation}\label{DeltaE}
\Delta E^{(M)}= \frac{-4 t^{M}}{U^{M-1}}
\sum_{\beta,\beta'}
\sum_{{\bf A}_{\rm f}^{(\beta,\beta')}}
 \frac{f_\beta f_{\beta'}\, {\rm sign}P_{\uparrow}\left({\bf A}_{\rm
       f}^{(\beta,\beta')}\right)\, {\rm sign}P_{\downarrow}\left({\bf
       A}_{\rm f}^{(\beta,\beta')}\right)}
  {(\epsilon_{\gamma_1}+d_{\gamma_1}W/U)
  (\epsilon_{\gamma_2}+d_{\gamma_2}W/U)\cdots
  (\epsilon_{\gamma_{M-1}}+d_{\gamma_{M-1}}W/U)}
\end{equation}
as sums over all forward hopping sequences ${\bf A}_{\rm f}^{(\beta,\beta')}$,
with an additional factor of two accounting for the corresponding backward 
sequences. We have defined 
$E_{\gamma_l}^{\rm W}-E_0^{\rm W}=Wd_{\gamma_l}$, and extracted the
dominant parametric dependence $U^{M-1}$ from the denominator.
All contributing sequences lead to a cyclic perturbation of the
$N/2$ operators corresponding to electrons with a given spin direction
in (\ref{state}), and since the weights $f_\beta$ are all
positive, the sign of the phase sensitivity 
at strong interaction is given by $(-1)^{N/2}$, as in the
non-interacting case \cite{selva00}.  

In order to study the effect of disorder on the phase sensitivity
at strong interaction ($U\gg W$), we expand (\ref{DeltaE}) in powers of $W/U$.
Up to second order, this yields
\begin{equation}\label{Wexpansion}
\Delta E^{(M)}\approx\frac{
  (-1)^{N/2}4t^{M}}{U^{M-1}}\sum_{\beta,\beta'}\sum\limits_{{\bf A}_{\rm f}^{(\beta,\beta')}}
\frac{f_\beta f_{\beta'}}{\prod_{l}\epsilon_{\gamma_l}}
\left(1-\frac{W}{U}\sum_{l}\frac{d_{\gamma_l}}{\epsilon_{\gamma_l}}
+\frac{W^2}{U^2}
\left(\sum_l\frac{d_{\gamma_l}^2}{\epsilon_{\gamma_l}^2}
+\sum_{l<m}\frac{d_{\gamma_l}d_{\gamma_m}}
                {\epsilon_{\gamma_l}\epsilon_{\gamma_m}}\right)\right)\, .
\end{equation}
Since the matrix elements of the antiferromagnetic Heisenberg
Hamiltonian are functions of $(W/U)^2$, the components $f_\beta$ of
its ground state must be even functions of $W/U$.
Up to second order, $f_\beta\approx f_\beta^{(0)}+
f_\beta^{(2)}W^2/U^2$.
The normalization condition $\sum_\beta f_\beta^2 =1$ implies  
$\sum_\beta f_\beta^{(0)} f_\beta^{(2)} =0$, and since
$f_\beta^{(0)}>0$ we must have positive and negative coefficients
$f_\beta^{(2)}$.
The dominating term in (\ref{Wexpansion}), corresponding to the clean 
limit $W\rightarrow 0$, 
exhibits an interaction-induced suppression of the persistent
current $\propto U(t/U)^M$. This suppression gets exponentially more 
pronounced as the system size increases. As in the spinless case
\cite{weinmann01}, this can be interpreted as an interaction-dependent 
localization length $\xi(U)=1/\ln(U/t)$. 

In contrast to the cases of lower filling or spinless fermions, where
the linear correction in $W$ decreases the phase sensitivity
\cite{selva00}, in the half-filled Hubbard-Anderson chain, the
first-order correction in $W/U$ vanishes exactly. 
We demonstrate this in the following way:
a given sequence of hops can be characterized by the subsequent positions  
of doubly occupied sites and empty sites. Starting from this, 
one can construct a second sequence by exchanging the positions of 
the doubly occupied and empty sites. 
These two sequences have the same numbers of doubly occupied sites
$\epsilon_{\alpha_l}$, and the coefficients $d_{\alpha_l}$ have
opposite sign. Their contributions to the first-order 
correction in $W/U$ cancel each other, and the sum over all 
sequences of the linear term in the 
expansion (\ref{Wexpansion}) vanishes. The same argument
applies to higher odd orders of the expansion, all of which vanish.

Thus, the second order term of (\ref{Wexpansion}) determines the
disorder dependence of the persistent current at $U\gg W$. 
The first component of its prefactor is a sum over $M-1$ positive 
quantities of order one, and can be estimated to 
$\sum_l (d_{\alpha_l}/\epsilon_{\alpha_l})^2 \lesssim M$ per sequence. 
The second component is a sum over
terms which are also of order one, but do not have a preferential
sign. In the sum over all sequences, this yields, assuming random 
signs, a correction which scales with the square root of the number of 
sequences $N_{\bf A}\gtrsim M!$, and can thus be neglected as compared to the 
first component which scales $\propto N_{\bf A}$. The same holds for
the second order term emerging from the expansion of the $f_\beta$. 
Further, the fourth order correction (not explicitly written in
Eq.~\ref{Wexpansion}) is dominated by a positive term yielding a
prefactor of the order $M^2/2$ per sequence.   
This leads to the important conclusion that {\it disorder
increases the phase sensitivity (and therewith the stiffness) and the 
persistent current of strongly interacting electrons}. Moreover, the 
relative importance of the corrections increases with the system size $M$.

The increase is specific to half filling. At other fillings, 
the symmetry between doubly occupied and empty sites
disappears, and the non-vanishing first order term decreases the 
phase sensitivity. 
This insight is consistent with the conclusion of Ref. \cite{denteneer01} that 
particle-hole symmetry breaking leads to a disorder-induced increase of 
the low-temperature conductivity in 
two dimensions.

In order to verify and extend the analytical result derived above,
we performed extensive numerical calculations for the examples of 
$N=10$ particles ($5\uparrow; 5\downarrow$) on $M=10$ sites and 
$N=20$ particles ($10\uparrow; 10\downarrow$) on $M=20$ sites, using the 
Density Matrix Renormalization Group algorithm
\cite{white93,peschel99}. This algorithm allows to compute the many-body 
ground-state energies $E(0)$ and $E(\pi)$ for the interacting disordered 
system with sufficient precision to evaluate 
$\Delta E= E(0)-E(\pi)$ \cite{ED}. 
For $N=10$ $(20)$, we found negative (positive) $\Delta E$ for all
values of $U$ and $W$, consistent
with the non-interacting case and the analytical result for very 
strong interaction $\propto (-1)^{N/2}$, in agreement with the result
of Ref.~\cite{selva00}. 
\begin{figure}[tb]
\includegraphics*[height=5cm,angle=0]{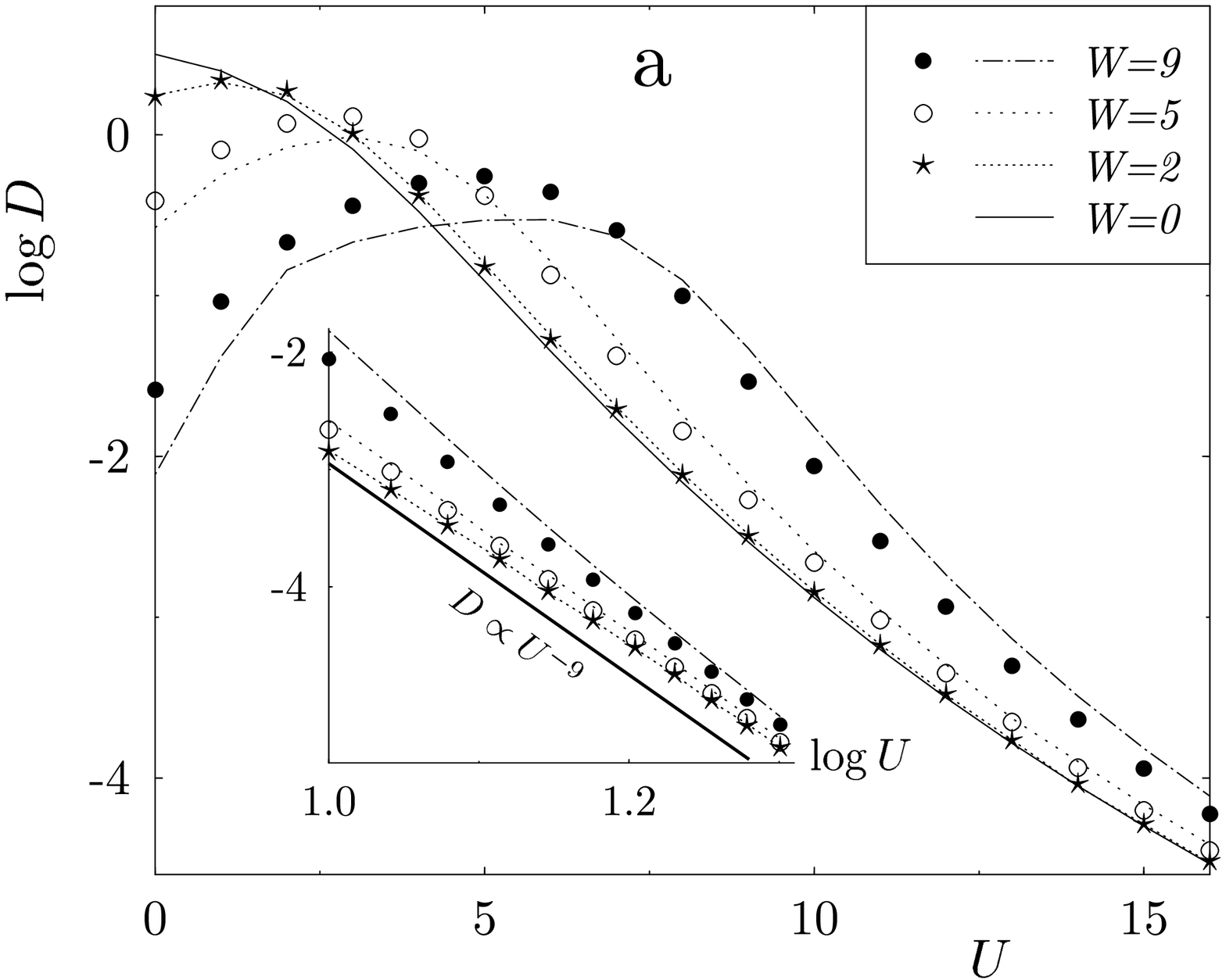}\hfill
\includegraphics*[height=5cm,angle=0]{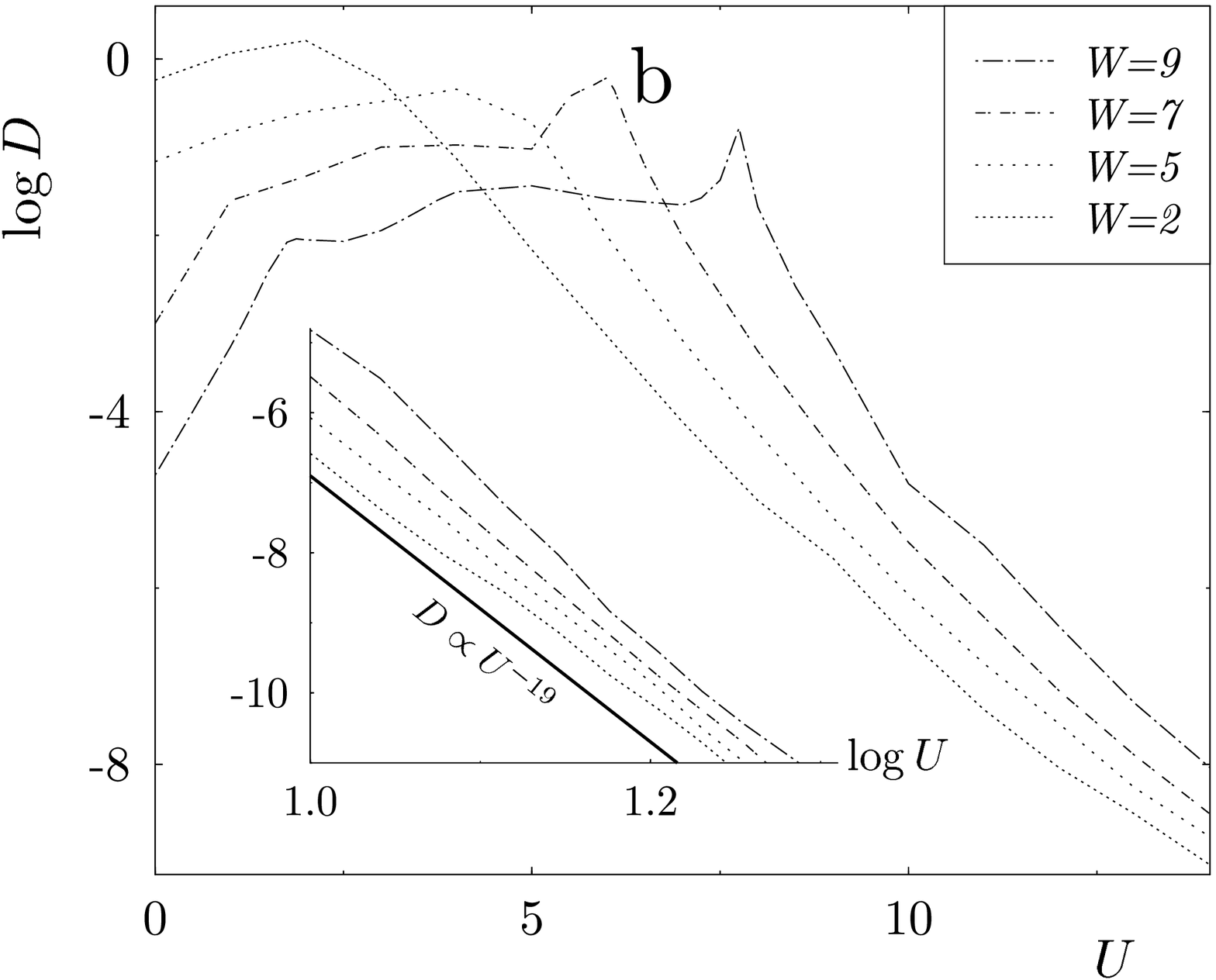}
\caption{\label{data} 
Interaction dependence of the stiffness for $M=N=10$ (a) and $M=N=20$ (b).
Lines represent $\log D$ as a function of the interaction strength $U$,
for different values of the disorder $W$, for an individual sample.
Symbols in (a) correspond to the ensemble average $\langle \log D \rangle$. 
The statistical errors are smaller than the symbol size.   
In the insets, strong-interaction data are plotted on log-log scale,
confirming the power law $D \sim U^{-(M-1)}$ (thick lines).}
\end{figure}

The disorder dependence of $|\Delta E|$ for $N=M=10$ particles  
is shown in Fig.~\ref{hops}~(c) for a typical individual sample (disorder
realization), at several large values of the interaction. The
numerical results are very well described by Eq.(\ref{Wexpansion}) for
small ratios of $W/U$. In order to have an agreement over a large
range of disorder (like in Fig.~\ref{hops}~(c)), we need to go to
fourth order in $W$. The parametric dependence $|\Delta
E|=(A_0/U^{9})(1+A_1(W/U)^2+A_2(W/U)^4)$ (solid lines) yields the values 
$A_0\approx 390000$, $A_1\approx 2.33$, and $A_2\approx 9.36$ 
(valid for all curves), consistent with the above presented estimations. 

In Fig.~\ref{data}, we show the interaction dependence of the
stiffness $D$ for $M=N=10$ (a) and $M=N=20$ (b) 
for typical individual samples (lines) and ensemble averages
(symbols), for different disorder values $W$, and for the clean case
$W=0$ (solid line). 
The ensemble averages (of $\log D$) are performed over 100 
different samples (disorder realizations).  
At $U=0$, the disorder leads to Anderson localization, and $D$ 
is strongly suppressed by increasing disorder (note the logarithmic scale). 
While the interaction always reduces the stiffness in clean rings, consistent
with a Luttinger liquid calculation \cite{loss92},
a weak repulsive interaction $U\lesssim t$ leads to an increase of the
stiffness when disorder is present. 
Such an increase was predicted from a renormalization group
approach \cite{giamarchi95}, away from half filling.
Here, at half filling, the increase becomes very pronounced as the 
disorder increases.

At large repulsive interaction $U\gg W$, in the Mott insulator limit, 
the behavior is radically different. The stiffness decreases 
strongly with the interaction, and, as shown in the insets of 
Fig.~\ref{data}, the numerical results agree with the predicted power 
law $\propto 1/U^{M-1}$. While such an agreement is already obtained
for individual samples, it persists for the average values. Moreover, and in
contrast to the Anderson insulator, {\it{the disorder increases the phase 
sensitivity of the Mott insulator}}. This enhancement can be very
strong and become larger than an order of magnitude. 
At intermediate interaction, 
between the two previous limits ($t\lesssim U\lesssim W$), the phase 
sensitivity exhibits a maximum which becomes broader with increasing 
disorder and the disorder-induced decrease of its height is much
weaker than for the Anderson insulator.

Our analysis over two sizes confirms that the disorder-induced
increase of the persistent currents becomes more pronounced 
(in relative terms) as the system size increases. In the case of
$N=M=10$, where the ensemble average can be readily performed, we 
observe a similar behavior for individual samples and the average. 
Noticeable differences only appear at very strong disorder, but the 
analytically predicted enhancement of the phase sensitivity is robust 
with respect to impurity average. For the large system size $M=N=20$, 
small peaks appear in the interaction-dependence of $D$, in individual
samples at strong disorder. This structure is less pronounced than in 
the case of spinless fermions in strong disorder $W=9$
\cite{schmitteckert98,weinmann01}, where sharp peaks appear at 
realization-dependent values of $U$. It is also important to notice
that the data of Fig.~\ref{data}~ exhibit, for all finite values of the
disorder, a pronounced increase of the stiffness at intermediate $U$ 
(with respect to the noninteracting case). This enhancement for the 
average can be larger than an order of magnitude, while for spinless
fermions a small maximum appears only at very strong disorder 
\cite{schmitteckert98,weinmann01}.

In conclusion, we have investigated the effect of disorder on the 
persistent current of interacting electrons in one-dimensional 
half-filled chains, taking into account the spin degree of freedom. 
Within a systematic perturbative expansion of the phase sensitivity,
we have unambiguously demonstrated that disorder increases the
persistent current in the presence of strong interactions, in the Mott 
insulator limit. This is in striking contrast to the non-interacting 
Anderson insulator limit, where disorder localizes the electrons and
suppresses the persistent current. Such an effect can be related to the 
disorder-induced reduction of the energy gap between the Mott
insulator ground state and excited states. Moreover, we have also
shown that, for intermediate interactions between the two strongly 
insulating Anderson and Mott limits ($t\lesssim U\lesssim W$), the 
stiffness exhibits a pronounced maximum, consistently with the results
away from half filling. Such an effect might be a precursor of an even 
stronger tendency towards interaction-induced delocalization in
disordered two-dimensional systems.

\acknowledgments
We thank T.\ Giamarchi, G.-L.\ Ingold, and J.-L.\ Pichard for useful 
discussions and F.\ Maingot for computer assistance. This work was 
supported in part by the European Union within the RTN program, and 
by the NSF under Grant N.\ PHYS 99-07949.


\begin{thebibliography}{99}
\bibitem{excitedS}
In principle, the excited states of $\cal S$ should
have been included in the perturbative expansion of Eq.(\ref{energyperturb}). 
However, sequences passing through these states and transferring one particle 
across the ring are of higher order than the ones kept.
\bibitem{ED}
We used eight finite-lattice iterations and kept up to 600 states per block.
\end{thebibliography}
\end{document}